\pdfoutput=1
\documentclass[12pt,a4paper]{article}
\usepackage{ifthen} % for conditional statements
\newboolean{pdflatex}
\setboolean{pdflatex}{true} % False for eps figures 

\newboolean{articletitles}
\setboolean{articletitles}{true} % False removes titles in references

\newboolean{uprightparticles}
\setboolean{uprightparticles}{false} %True for upright particle symbols

\def\paperauthors{LHCb collaboration} % Leave as is for PAPER, CONF and FIGURE
\def\paperasciititle{Template for writing LHCb papers} % Set ASCII title here !! MAKE sure it's only ASCII characters !! 
\def\paperkeywords{{High Energy Physics}, {LHCb}} % Comma separated list
\def\papercopyright{\the\year\ CERN for the benefit of the LHCb collaboration} % new since 9/Apr/2018
\def\paperlicence{CC BY 4.0 licence}
\def\paperlicenceurl{https://creativecommons.org/licenses/by/4.0/}

\usepackage[top=1in, bottom=1.25in, left=1in, right=1in]{geometry}

\usepackage{microtype}
\usepackage{lineno}  % for line numbering during review
\usepackage{xspace} % To avoid problems with missing or double spaces after
\usepackage{caption} %these three command get the figure and table captions automatically small

\usepackage{graphicx}  % to include figures (can also use other packages)
\usepackage{color}
\usepackage{colortbl}
\graphicspath{{./figs/}} % Make Latex search fig subdir for figures
\usepackage{amsmath} % Adds a large collection of math symbols
\usepackage{amssymb}
\usepackage{amsthm}
\usepackage{amsfonts}
\usepackage{upgreek} % Adds in support for greek letters in roman typeset

\newcommand*\patchAmsMathEnvironmentForLineno[1]{%
\expandafter\let\csname old#1\expandafter\endcsname\csname #1\endcsname
\expandafter\let\csname oldend#1\expandafter\endcsname\csname
end#1\endcsname
 \renewenvironment{#1}%
   {\linenomath\csname old#1\endcsname}%
   {\csname oldend#1\endcsname\endlinenomath}%
}
\newcommand*\patchBothAmsMathEnvironmentsForLineno[1]{%
  \patchAmsMathEnvironmentForLineno{#1}%
  \patchAmsMathEnvironmentForLineno{#1*}%
}
\AtBeginDocument{%
\patchBothAmsMathEnvironmentsForLineno{equation}%
\patchBothAmsMathEnvironmentsForLineno{align}%
\patchBothAmsMathEnvironmentsForLineno{flalign}%
\patchBothAmsMathEnvironmentsForLineno{alignat}%
\patchBothAmsMathEnvironmentsForLineno{gather}%
\patchBothAmsMathEnvironmentsForLineno{multline}%
\patchBothAmsMathEnvironmentsForLineno{eqnarray}%
}

\usepackage{hyperxmp}

\usepackage[pdftex,
            pdfauthor={\paperauthors},
            pdftitle={\paperasciititle},
            pdfkeywords={\paperkeywords},
            pdfcopyright={Copyright (C) \papercopyright},
            pdflicenseurl={\paperlicenceurl}]{hyperref}
\usepackage[colorinlistoftodos,textsize=scriptsize]{todonotes}

\usepackage[bottom,flushmargin,hang,multiple]{footmisc}

\usepackage[all]{hypcap} % Internal hyperlinks to floats.

\usepackage{cite} % Allows for ranges in citations
\usepackage{mciteplus}
\usepackage{footmisc}

\usepackage{mciteplus}
\usepackage{footmisc}
\usepackage{longtable} % only for template; not usually to be used in PAPERs
\usepackage{booktabs} % for complicated table
\usepackage{rotating} % Rotating table
\usepackage{lscape}
\usepackage{units}
\usepackage{overpic}
\usepackage{rotating} 
\usepackage{subfig}
\usepackage{pbox}
\usepackage{tabularx}
\usepackage{multirow}
\usepackage{multicol}
\usepackage{listings}
\usepackage{color}
\usepackage[none]{hyphenat}
\usepackage{amsmath}
\usepackage{xspace} 
\usepackage{upgreek}

\def\lhcb   {\mbox{LHCb}\xspace}

\def\MagUp {\mbox{\em Mag\kern -0.05em Up}\xspace}

\ifthenelse{\boolean{uprightparticles}}%
{

 \def\Ppi         {\ensuremath{\uppi}\xspace}

 \def\Pphi        {\ensuremath{\upphi}\xspace}

 \def\Ppsi        {\ensuremath{\uppsi}\xspace}

 \def\PDelta      {\ensuremath{\Delta}\xspace}                 
 \def\PXi         {\ensuremath{\Xi}\xspace}                 
 \def\PLambda     {\ensuremath{\Lambda}\xspace}                 
 \def\PSigma      {\ensuremath{\Sigma}\xspace}                 
 \def\POmega      {\ensuremath{\Omega}\xspace}                 
 \def\PUpsilon    {\ensuremath{\Upsilon}\xspace}
 \let\oldPi\Pi
 \def\PPi         {\ensuremath{\oldPi}\xspace}

 \def\PB      {\ensuremath{\mathrm{B}}\xspace}                 
                  
 \def\PD      {\ensuremath{\mathrm{D}}\xspace}

 \def\PJ      {\ensuremath{\mathrm{J}}\xspace}                 
 \def\PK      {\ensuremath{\mathrm{K}}\xspace}

 \def\Pb      {\ensuremath{\mathrm{b}}\xspace}

 \def\Pi      {\ensuremath{\mathrm{i}}\xspace}

 \def\Ps      {\ensuremath{\mathrm{s}}\xspace}

 \def\thebaroffset{0.0em}
}
{

 \def\Ppi         {\ensuremath{\pi}\xspace}

 \def\Pphi        {\ensuremath{\phi}\xspace}

 \def\Ppsi        {\ensuremath{\psi}\xspace}                 
                  
 \mathchardef\PDelta="7101
 \mathchardef\PXi="7104
 \mathchardef\PLambda="7103
 \mathchardef\PSigma="7106
 \mathchardef\POmega="710A
 \mathchardef\PUpsilon="7107
 \mathchardef\PPi="7105
                  
 \def\PB      {\ensuremath{B}\xspace}                 
                  
 \def\PD      {\ensuremath{D}\xspace}

 \def\PJ      {\ensuremath{J}\xspace}                 
 \def\PK      {\ensuremath{K}\xspace}

 \def\Pb      {\ensuremath{b}\xspace}

 \def\Pi      {\ensuremath{i}\xspace}

 \def\Ps      {\ensuremath{s}\xspace}

 \def\thebaroffset{0.18em}
}
\newcommand{\offsetoverline}[2][\thebaroffset]{\kern #1\overline{\kern -#1 #2}}%

\makeatletter
\ifcase \@ptsize \relax% 10pt
  \newcommand{\miniscule}{\@setfontsize\miniscule{4}{5}}% \tiny: 5/6
\or% 11pt
  \newcommand{\miniscule}{\@setfontsize\miniscule{5}{6}}% \tiny: 6/7
\or% 12pt
  \newcommand{\miniscule}{\@setfontsize\miniscule{5}{6}}% \tiny: 6/7
\fi
\makeatother

\DeclareRobustCommand{\optbar}[1]{\shortstack{{\miniscule (\rule[.5ex]{1.25em}{.18mm})}
  \\ [-.7ex] $#1$}}

\def\squark    {{\ensuremath{\Ps}}\xspace}
\def\squarkbar {{\ensuremath{\overline \squark}}\xspace}

\def\bquark    {{\ensuremath{\Pb}}\xspace}

\def\pion   {{\ensuremath{\Ppi}}\xspace}

\def\pip    {{\ensuremath{\pion^+}}\xspace}
\def\pim    {{\ensuremath{\pion^-}}\xspace}

\def\kaon    {{\ensuremath{\PK}}\xspace}
\def\KorKbar {\kern \thebaroffset\optbar{\kern -\thebaroffset \PK}{}\xspace}

\def\Kp      {{\ensuremath{\kaon^+}}\xspace}
\def\Km      {{\ensuremath{\kaon^-}}\xspace}

\newcommand{\phiz}{\ensuremath{\Pphi}\xspace}

\def\D       {{\ensuremath{\PD}}\xspace}

\def\DorDbar {\kern \thebaroffset\optbar{\kern -\thebaroffset \PD}\xspace}

\def\Dp      {{\ensuremath{\D^+}}\xspace}
\def\Dm      {{\ensuremath{\D^-}}\xspace}

\def\DpDm    {\ensuremath{\Dp {\kern -0.16em \Dm}}\xspace}

\def\B       {{\ensuremath{\PB}}\xspace}
\def\Bbar    {{\ensuremath{\offsetoverline{\PB}}}\xspace}

\def\BorBbar {\kern \thebaroffset\optbar{\kern -\thebaroffset \PB}\xspace}

\def\Bd      {{\ensuremath{\B^0}}\xspace}

\def\BdorBdbar {\kern \thebaroffset\optbar{\kern -\thebaroffset \Bd}\xspace}

\def\Bs      {{\ensuremath{\B^0_\squark}}\xspace}
\def\Bsb     {{\ensuremath{\Bbar{}^0_\squark}}\xspace}
\def\BsorBsbar {\kern \thebaroffset\optbar{\kern -\thebaroffset \Bs}\xspace}

\def\jpsi     {{\ensuremath{{\PJ\mskip -3mu/\mskip -2mu\Ppsi}}}\xspace}

\def\Y#1S{\ensuremath{\PUpsilon{(#1S)}}\xspace}

\def\Lz          {{\ensuremath{\PLambda}}\xspace}

\def\LorLbar     {\kern \thebaroffset\optbar{\kern -\thebaroffset \PLambda}\xspace}

\def\Lb           {{\ensuremath{\Lz^0_\bquark}}\xspace}

\newcommand{\decay}[2]{\ensuremath{#1\!\to #2}\xspace} 

\def\to                 {\ensuremath{\rightarrow}\xspace}

\newcommand{\lhcborcid}[1]{\href{https://orcid.org/#1}{\hspace*{0.1em}\raisebox{-0.45ex}{\includegraphics[width=1em]{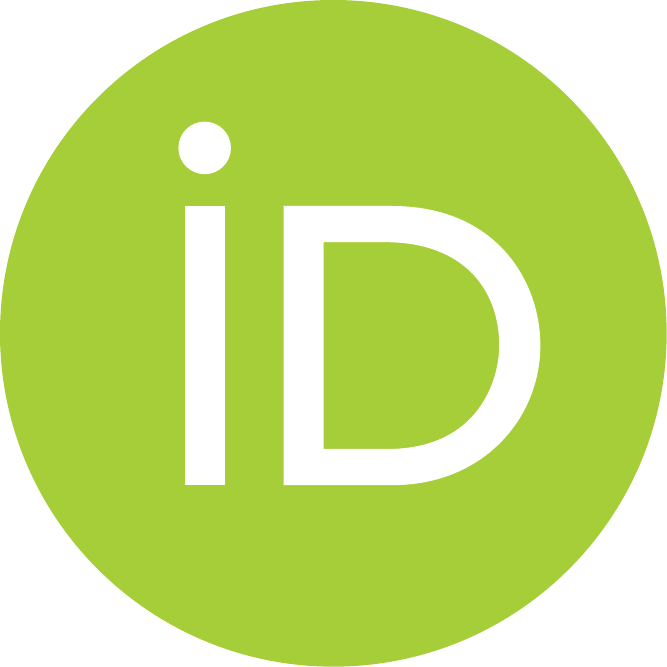}}}}
\newcommand{\Azero}{{\ensuremath{A_{0}}}\xspace}

\newcommand{\Aperp}{{\ensuremath{A_{\perp}}}\xspace}
\newcommand{\AS}{{\ensuremath{A_{S}}}\xspace}

\newcommand{\delzero}{{\ensuremath{\delta_{0}}}\xspace}
\newcommand{\delpara}{{\ensuremath{\delta_{\parallel}}}\xspace}
\newcommand{\delperp}{{\ensuremath{\delta_{\perp}}}\xspace}

\newcommand{\lamzero}{{\ensuremath{\lambda_{0}}}\xspace}
\newcommand{\lampara}{{\ensuremath{\lambda_{\parallel}}}\xspace}
\newcommand{\lamperp}{{\ensuremath{\lambda_{\perp}}}\xspace}

\newcommand{\gevt}{\ensuremath{\mathrm{Ge\kern -0.1em V}}\xspace}

\newcommand{\AAv}[1][\Ai]{#1^{\text{CP}}}

\newcommand{\Dms}{\Delta m_\text{s}}

\newcommand{\cDGs}{\cosh\!\left(\frac{\DGs}{2}t\right)}
\newcommand{\sDGs}{\sinh\!\left(\frac{\DGs}{2}t\right)}
\newcommand{\cDms}{\cos\!\left(\Dms\, t\right)}
\newcommand{\sDms}{\sin\!\left(\Dms\, t\right)}

\newcommand{\magSAv}[1][]{|\AAv[\AS]|}
\newcommand{\magSAvSq}[1][]{\magSAv^2}

\newcommand{\phipa}{\ensuremath{\phi_{s,\parallel}}\xspace}
\newcommand{\phipe}{\ensuremath{\phi_{s,\perp}}\xspace}

\def\CP                {{\ensuremath{C\!P}}\xspace}

\newcommand{\DGs}{{\ensuremath{\Delta\Gamma_{\squark}}}\xspace}

\def\AT#1     {\ensuremath{A_{\mathrm{T}}^{#1}}\xspace}           % 2

\def\C#1      {\ensuremath{\mathcal{C}_{#1}}\xspace}                       % 9
\def\Cp#1     {\ensuremath{\mathcal{C}_{#1}^{'}}\xspace}                    % 7
\def\Ceff#1   {\ensuremath{\mathcal{C}_{#1}^{\mathrm{(eff)}}}\xspace}        % 9  
\def\Cpeff#1  {\ensuremath{\mathcal{C}_{#1}^{'\mathrm{(eff)}}}\xspace}       % 7
\def\Ope#1    {\ensuremath{\mathcal{O}_{#1}}\xspace}                       % 2
\def\Opep#1   {\ensuremath{\mathcal{O}_{#1}^{'}}\xspace}                    % 7

\newcommand{\cbracket}[1]{\ensuremath{\left[#1\right]}} % [x]
\newcommand{\aunit}[1]{\ensuremath{\text{\,#1}}}       

\newcommand{\tev}{\aunit{Te\kern -0.1em V}\xspace}
\newcommand{\gev}{\aunit{Ge\kern -0.1em V}\xspace}
\newcommand{\mev}{\aunit{Me\kern -0.1em V}\xspace}
\newcommand{\kev}{\aunit{ke\kern -0.1em V}\xspace}
\newcommand{\ev}{\aunit{e\kern -0.1em V}\xspace}
 
\newcommand{\mevc}{\ensuremath{\aunit{Me\kern -0.1em V\!/}c}\xspace}
\newcommand{\gevc}{\ensuremath{\aunit{Ge\kern -0.1em V\!/}c}\xspace}
\newcommand{\mevcc}{\ensuremath{\aunit{Me\kern -0.1em V\!/}c^2}\xspace}
\newcommand{\gevcc}{\ensuremath{\aunit{Ge\kern -0.1em V\!/}c^2}\xspace}
\newcommand{\gevgevcc}{\ensuremath{\gev^2\!/c^2}\xspace} % for \pt^2 in CEP
\def\fb   {\ensuremath{\aunit{fb}}\xspace}
\def\invfb   {\ensuremath{\fb^{-1}}\xspace}

\def\ps   {\ensuremath{\aunit{ps}}\xspace}
\def\fs   {\aunit{fs}}

\newcommand{\stat}{\aunit{(stat)}\xspace}
\newcommand{\syst}{\aunit{(syst)}\xspace}

\def\gsim{{~\raise.15em\hbox{$>$}\kern-.85em
          \lower.35em\hbox{$\sim$}~}\xspace}
\def\lsim{{~\raise.15em\hbox{$<$}\kern-.85em
          \lower.35em\hbox{$\sim$}~}\xspace}

\newcommand{\abs}[1]{\ensuremath{\left|#1\right|}} % {x}

\def\sPlot{\mbox{\em sPlot}\xspace}

\def\pt         {\ensuremath{p_{\mathrm{T}}}\xspace}

\def\rad{\aunit{rad}\xspace}

\def\tell1  {TELL1\xspace}
\def\ukl1   {UKL1\xspace}

\newcommand{\phisss}{{\ensuremath{\phi_{\squark}^{\squark\squarkbar\squark}}}\xspace}

\begin{document}

%%%%%%%%%%%%%%%%%%%%%%%%%
%%%%% Title     %%%%%%%%%
%%%%%%%%%%%%%%%%%%%%%%%%%
\renewcommand{\thefootnote}{\fnsymbol{footnote}}
\setcounter{footnote}{1}

% %%%%%%% CHOOSE TITLE PAGE--------
%\onecolumn
\begin{titlepage}
\pagenumbering{roman}

% Header ---------------------------------------------------
\vspace*{-1.5cm}
\centerline{\large EUROPEAN ORGANIZATION FOR NUCLEAR RESEARCH (CERN)}
\vspace*{1.5cm}
\noindent
\begin{tabular*}{\linewidth}{lc@{\extracolsep{\fill}}r@{\extracolsep{0pt}}}
\ifthenelse{\boolean{pdflatex}}% Logo format choice
{\vspace*{-1.5cm}\mbox{\!\!\!\includegraphics[width=.14\textwidth]{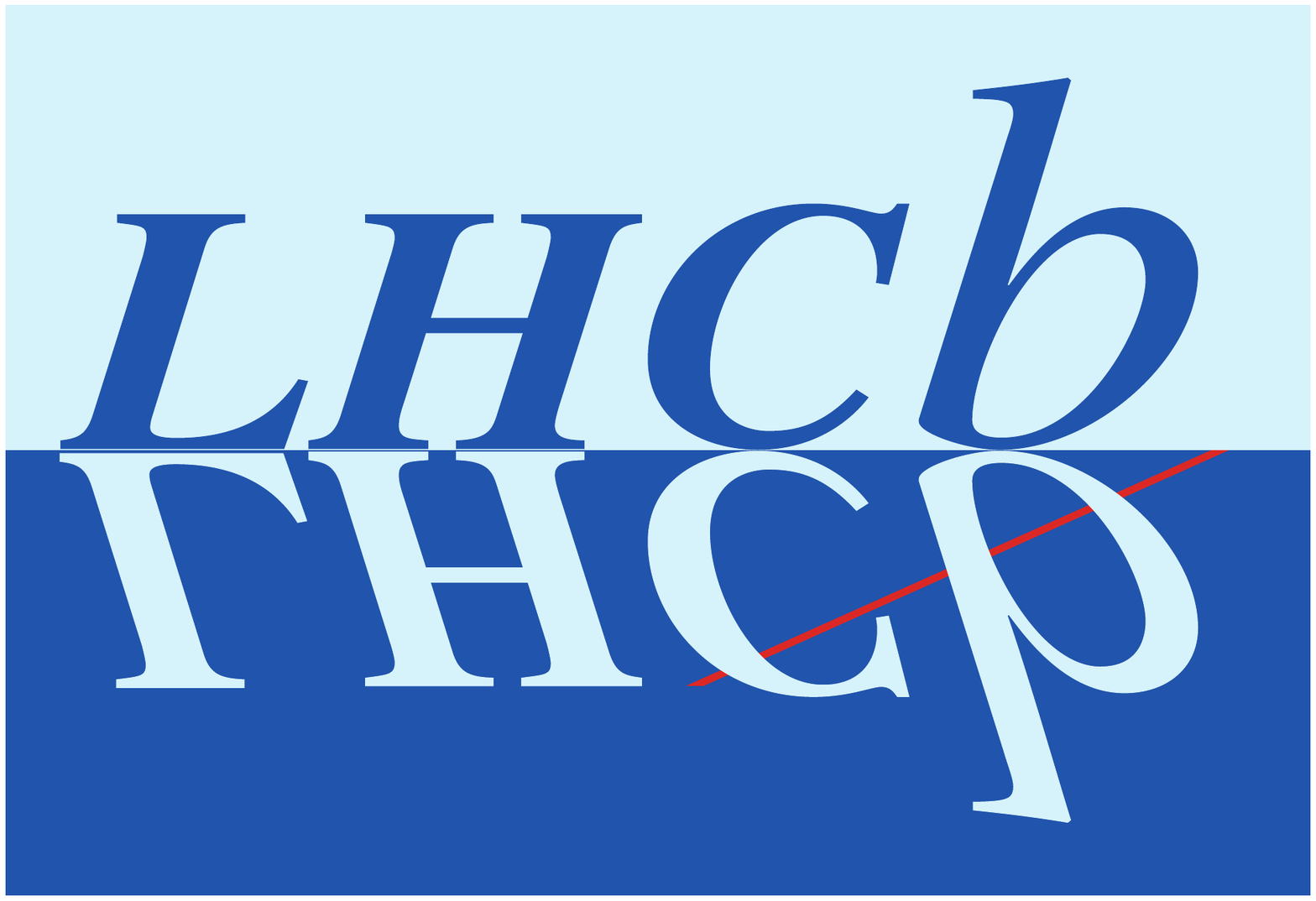}} & &}%
{\vspace*{-1.2cm}\mbox{\!\!\!\includegraphics[width=.12\textwidth]{lhcb-logo.eps}} & &}%
\\
 & & CERN-EP-2023-042 \\  % ID 
 & & LHCb-PAPER-2023-001 \\  % ID 
 & & 20 April 2023\\%\today \\ % Date - Can also hardwire e.g.: 23 March 2010
 & & \\
% not in paper \hline
\end{tabular*}

\vspace*{2.0cm}

% Title --------------------------------------------------
{\normalfont\bfseries\boldmath\huge
\begin{center}
\noindent
% DO NOT EDIT HERE. Instead edit macro in main.tex to keep metadata correct
Precision measurement of  \CP violation in the penguin-mediated decay $\Bs\to \phi \phi$
%  \papertitle 
\end{center}
}

% Authors -------------------------------------------------

\vspace*{1.0cm}
\begin{center}
 \paperauthors\footnote{Authors are listed at the end of this paper.}

\end {center}
%\vspace{\fill}

% Abstract -----------------------------------------------
\begin{abstract}
  \noindent
   \noindent
  \sloppy{}
 A flavor-tagged time-dependent angular analysis of the  decay \mbox{$\Bs\to\phi\phi$}  is performed 
 using  $pp$ collision data
 collected by the LHCb experiment at
% at $\sqrt{s}=13$ TeV, 
 the center-of-mass energy of {\mbox 13\tev}, 
corresponding to an integrated luminosity of {\mbox 6\invfb}.
 The \CP-violating phase and direct \CP-violation parameter  are measured to be  \mbox{$\phisss = -0.042 \pm 0.075  \pm 0.009 $}\rad and 
 \mbox{$\abs{\lambda}=1.004\pm 0.030  \pm 0.009 $}, 
 respectively,   
 assuming the same values for all polarization states of the $\phi\phi$ system. In these results, 
 the first uncertainties
are statistical and the second  systematic.
These parameters are also determined separately for each polarization state, showing  no evidence for polarization dependence.
 The results are combined with  previous LHCb measurements using  $pp$ collisions at center-of-mass energies of 7 and {\mbox 8\tev}, yielding $\phisss  = -0.074 \pm 0.069 $\rad and $\abs{\lambda}=1.009 \pm 0.030$. 
 This is the most precise study of time-dependent  \CP violation in a penguin-dominated  $B$ meson decay.
  The results   are consistent with \CP symmetry and with the Standard Model predictions. 
\end{abstract}
\vspace{\fill}
\begin{center}
Published in Phys. Rev. Lett. 131 (2023) 171802
\end{center}

%\vspace{\fill}

{\footnotesize 
% Edit macro in main.tex to keep metadata correct
\centerline{\copyright~\papercopyright. \href{\paperlicenceurl}{\paperlicence}.}}
\vspace*{2mm}

\end{titlepage}

%%%%%%%%%%%%%%%%%%%%%%%%%%%%%%%%
%%%%%  EOD OF TITLE PAGE  %%%%%%
%%%%%%%%%%%%%%%%%%%%%%%%%%%%%%%%

%  empty page follows the title page ----
\newpage
\setcounter{page}{2}
\mbox{~}
%\newpage
%
%% Author List ----------------------------
%%  You need to get a new author list!
%\input{LHCb_authorlist.tex}
%

\cleardoublepage
%\twocolumn
% %%%%%%%%%%%%% ---------

\renewcommand{\thefootnote}{\arabic{footnote}}
\setcounter{footnote}{0}

%%%%%%%%%%%%%%%%%%%%%%%%%%%%%%%%
%%%%%  Table of Content   %%%%%%
%%%%%%%%%%%%%%%%%%%%%%%%%%%%%%%%
%%%% Uncomment if desired
%\tableofcontents
\cleardoublepage

%%%%%%%%%%%%%%%%%%%%%%%%%
%%%%% Main text %%%%%%%%%
%%%%%%%%%%%%%%%%%%%%%%%%%

\pagestyle{plain} % restore page numbers for the main text
\setcounter{page}{1}
\pagenumbering{arabic}

%% Uncomment during review phase. 
%% Comment before a final submission.
%\linenumbers

%% This is the main body
%% It is useful to have a single file so comemnts are not missed in overleaf.
%\input{body}

% Do not include this in any draft (just for information in the template)
% Comment this in for paper drafts; do not include this in analysis note, conference and figure reports
%\input{acknowledgements}

Probing the nature of \CP violation is central to understanding  how the matter-dominated universe came to existence. Precision studies of time-dependent \CP asymmetries in $b\to c\Bar{c} s$ decays at the $e^{+}e^{-}$ $B$ factories~\cite{Belle-II:2010dht,Aubert:2003ii} helped establish the CKM mechanism~\cite{Cabibbo:1963yz,Kobayashi:1973fv} as the dominant source of \CP violation in the  Standard Model (SM). However, the amount of \CP violation provided by this mechanism  is insufficient to explain the observed matter-antimatter asymmetry~\cite{Huet:1994jb}.  Flavor-changing neutral current (FCNC)
decays of $B$ mesons are highly sensitive to new physics contributions  entering via loop (penguin) processes and 
 provide excellent opportunities to  reveal new sources of \CP violation. 
 
 The penguin-dominated  decay $\Bs\to\phi\phi$, which proceeds via a $b\to s\Bar{s}s$ transition,     is a benchmark channel to study  \CP violation in FCNC decays at the \lhcb  experiment\footnote{Charge-conjugation processes are implied throughout this Letter.}.
 Time-dependent \CP violation in this decay,  arising from the interference between the direct decay and the  decay after \Bs  mixing,  is characterized by the \CP-violating  phase $\phisss$  and 
 the parameter  $\abs{\lambda}$, which is related to direct \CP violation.
In the SM,  the phase  $\phisss$ is expected to be very close to zero due to a cancellation of the mixing and decay weak phases~\cite{Raidal:2002ph,  Beneke:2006hg, Bartsch:2008ps, Cheng:2009mu, Wang:2017rmh}, and the parameter $\abs{\lambda}$ is expected to be close to unity, indicating vanishing direct \CP asymmetry~\cite{Beneke:2006hg, Bartsch:2008ps, Cheng:2009mu, Wang:2017rmh, Yan:2018fif}. However, new physics contributions in the penguin decay
or the \Bs mixing could  significantly alter the values of  $\phisss$ and $\abs{\lambda}$~\cite{Zhang:2000ic, Datta:2012ky, Bhattacharya:2013sga, Kapoor:2023txh}.

In addition, the \mbox{$\phi\phi$} system in this decay is produced in three  linear polarization states, and
the effects of new physics may be polarization dependent~\cite{Bhattacharya:2013sga, Kapoor:2023txh}.
\textcolor{black} {
This is in contrast to the \mbox{$\Bs\to\jpsi\phi$} decay, where potential new physics would mainly affect the mixing process and the polarization dependence is expected to be less pronounced, since the decay amplitude is dominated by a  tree-level $b\to c\Bar{c}s$  diagram. With no hint of new physics found in the $\Bs$ mixing,  this study of \CP violation in  $\Bs \to \phi\phi$ decays  aims to probe new physics in $b \to s$ penguin processes.}

The \lhcb  collaboration has previously studied \CP violation in the decay \mbox{$\Bs\to\phi\phi$}~\cite{LHCb-PAPER-2013-007, LHCb-PAPER-2014-026, LHCb-PAPER-2019-019}. The  most precise measurements, 
$\phisss=-0.073\pm 0.115  \stat \pm 0.027   \syst $\rad  and $\abs{\lambda}= 0.99 \pm 0.05   \stat\pm  0.01   \syst$~\cite{LHCb-PAPER-2019-019}, are    based on a data sample collected before 2017, corresponding to an integrated luminosity of 5\invfb. 
Due to the limited size of the data sample, a full polarization-dependent analysis was not performed.   
Instead,  the \CP-violating phases for the parallel and perpendicular polarization states were measured separately, with the phase of the longitudinal polarization fixed to zero. No evidence for polarization-dependent \CP violation  was observed.

This Letter reports an updated  measurement of the \CP-violation parameters in  \mbox{$\Bs\to\phi\phi$ }decays using  the full
data sample of  $pp$ collisions  collected with the \lhcb  detector at a center-of-mass energy of {\mbox 13\tev} in 2015-2018 (Run 2), 
corresponding to an integrated luminosity of {\mbox 6\invfb}. The results are then combined with the \lhcb   result based on the data collected at  7 and {\mbox 8\tev} in 2011 and 2012 (Run 1)~\cite{LHCb-PAPER-2014-026}, corresponding to an integrated 
luminosity of 3\invfb.  Apart from the increased data sample size, this analysis also benefits from recent software improvements
including new reconstruction algorithms for the vertex detectors that lead to a better decay-time resolution, and
 optimized flavor-tagging algorithms that  increase the  effective tagging efficiency.
The  flavor-tagged time-dependent angular analysis  reported here largely follows the procedure described in Refs.~\cite{LHCb-PAPER-2019-019}. New features related to the calibration of flavor
tagging and decay-time resolution, and the modelling of  decay-time acceptance are described in detail below.  The increased sample size enables the \CP-violation parameters in the decay \mbox{$\Bs\to\phi\phi$} to be measured independently for all polarization states for the first time.

%%introduction of LHCb detector

The \lhcb  detector
is a single-arm forward spectrometer covering the pseudorapidity range \mbox{$2< \eta < 5$} and is described in detail in Refs.~\cite{LHCb-DP-2008-001,LHCb-DP-2014-002}.
The simulated events used in this analysis are produced with the software described in Refs.~\cite{Sjostrand:2007gs,Sjostrand:2006za,LHCb-PROC-2010-056,Agostinelli:2002hh,Allison:2006ve,LHCb-PROC-2011-006,davidson2015photos}.
The online event selection is performed by a 
trigger~\cite{LHCb_trigger},
which consists of a hardware trigger followed by a two-stage software trigger. For this analysis, the hardware trigger   selects hadron or muon candidates with high transverse momentum (\pt). In the software trigger, \mbox{$\Bs\to\phi\phi$}   candidates are selected
either by identifying events containing a pair of oppositely
charged kaons with an invariant mass close to that of the $\phi$
meson, or by using a three-body topological $b$-hadron trigger~\cite{LHCb-PROC-2015-018}.

In the offline selection, \mbox{$\Bs\to\phi\phi$} candidates, with the $\phi$ mesons decaying to a $\Kp\Km$ pair, are reconstructed by combining  four high-quality charged tracks compatible with originating from the same vertex and identified as kaons. 
All  tracks are required to have a \pt above 0.4 \gevc. 
Two  $\Kp\Km$ pairs, each with an invariant mass  within 20 $\mevcc$ of the known mass of the $\phi$ meson~\cite{PDG2022}, are combined to form \mbox{$\Bs\to\phi\phi$ }candidates and the product of their \pt is 
required to be above 1.2 \gevgevcc.  
The decay time of the $\Bs$ candidates must  be larger than  0.3\ps. The $\Kp\Km\Kp\Km$  invariant mass, $m(\Kp\Km\Kp\Km)$, must fall in the interval [5000, 5800] $\mevcc$.
 Potential contamination from \mbox{$\Bd\to\phi K^{*0}$} decays is largely suppressed by  
 removing    candidates that have a $\Kp\Km$
 invariant mass within 30
 $\mevcc$ of the known  $K^{*}(892)^{0}$  mass~\cite{PDG2022} when the pion mass is assigned to a  kaon candidate.

 A multilayer perceptron (MLP) classifier~\cite{Hocker:2007ht,MLP2001}  is used to further suppress  background and improve the signal significance. The vertex fit $\chi^{2}$ per degree of freedom,  \pt  and $\eta$ of the \Bs  candidate, the cosine of the angle between the \Bs momentum and its flight direction,  
 as well as  the \pt and $\eta$ of the kaon  and $\phi$  candidates are used as discriminating variables.  The MLP classifier is trained using 
 a sample of simulated  \mbox{$\Bs\to \phi \phi $} decays representing the  signal  and a background data sample consisting of  $\Bs\to \phi \phi $ candidates 
with an invariant mass outside a $\pm120$\mevcc window around  the known  \Bs  mass~\cite{PDG2022}. 
 The $k$-fold  method~\cite{Blum1999BeatingTH} with $k=8$ is employed in the training of the MLP classifier. 
 The lower threshold  on the MLP output is chosen by maximizing the figure-of-merit defined as $N_{s}/\sqrt{N_{s}+N_{b}}$, where $N_s$ is the expected signal yield in the signal region \mbox{[5322, 5412] $\mevcc$ } and $N_b$ is the expected  background yield in this region estimated by interpolating the number of candidates observed in the sideband regions \mbox{[5100, 5160] $\mevcc$} and \mbox{[5487, 5547] $\mevcc$}. 
Roughly $3\%$ of the events contain more than one candidate. In these cases one candidate is chosen at random, and the influence of this choice is found to be negligible.

\begin{figure}[tb]
\begin{center}
    \includegraphics[width=0.45\linewidth]{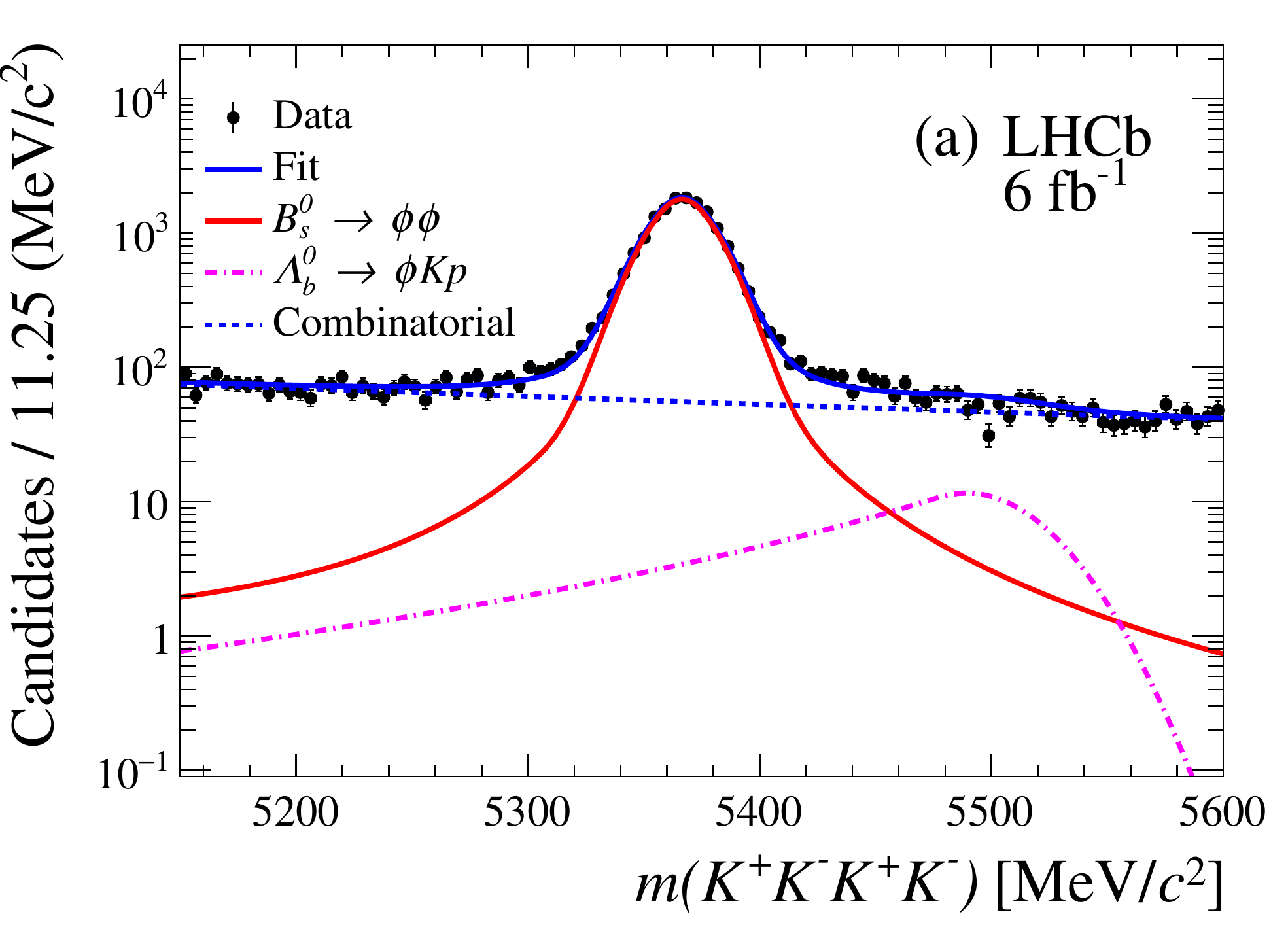} 
   \includegraphics[width=0.45\linewidth]{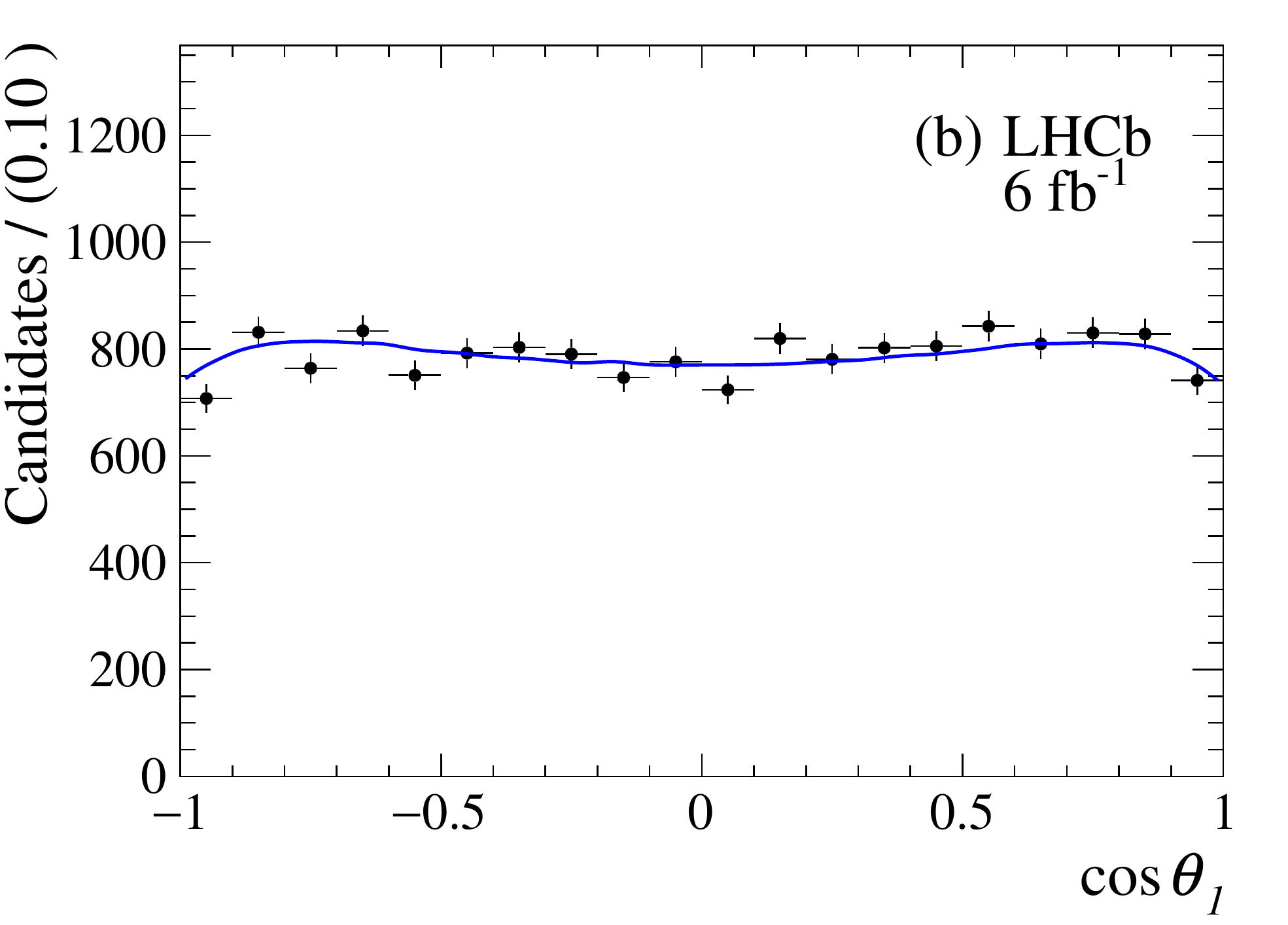}\\
   \includegraphics[width=0.45\linewidth]{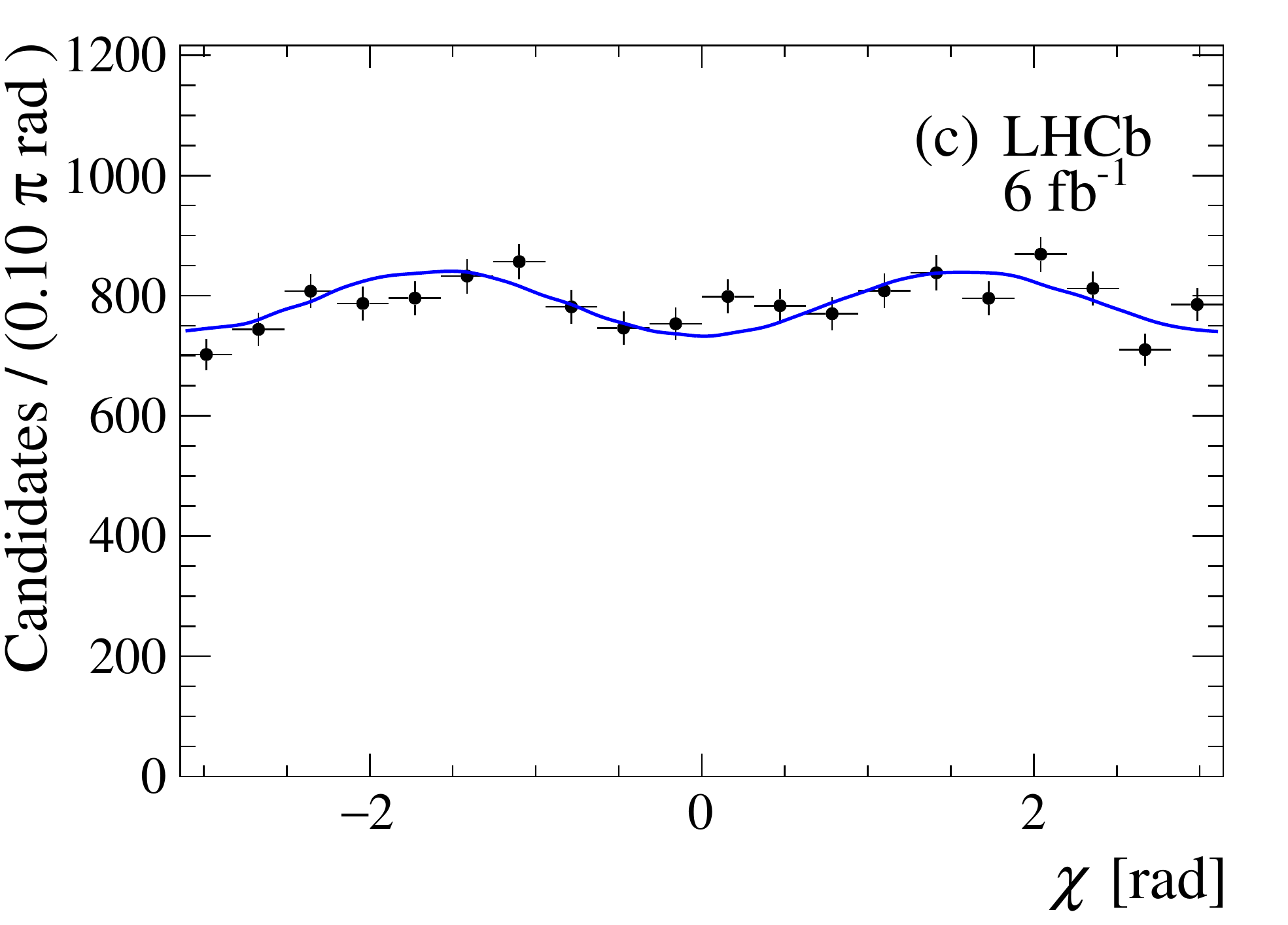}
   \includegraphics[width=0.45\linewidth]{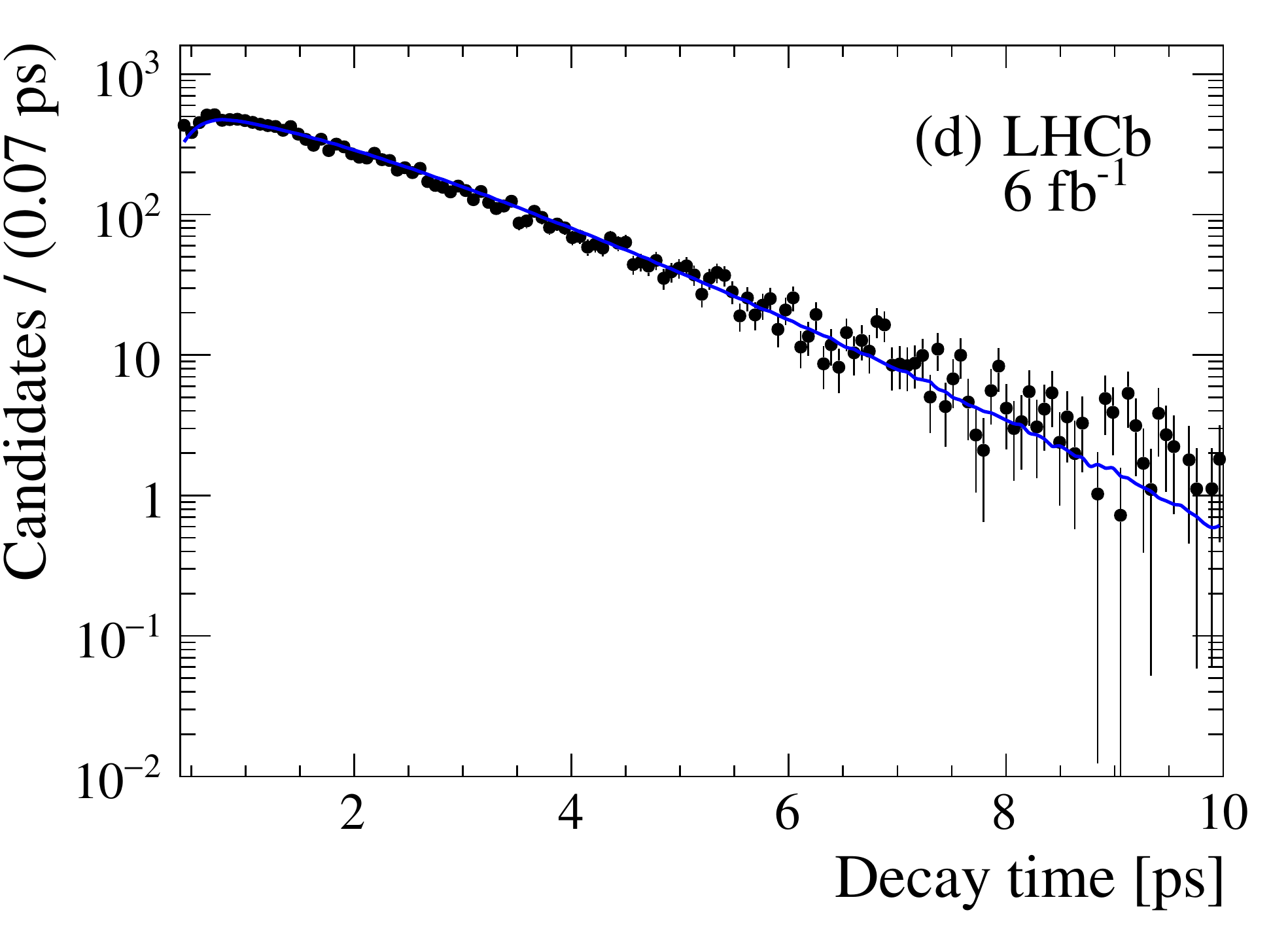}
  \vspace*{-0.5cm}
\end{center}
\caption{
    \textcolor{black}{(a)  Mass distribution of  the $\Bs\to\phi\phi$ candidates, superimposed by the fit projections.}
   (b-d) Background-subtracted distributions of  angular variables ($\cos\theta_{1}$ and $\chi$) and decay time, superimposed by the fit projections. 
 }

\label{fig_bs_finfit}
\end{figure}

Selected $\Bs\to\phi\phi$ candidates in the mass range [5150, 5600] $\mevcc$ are used in subsequent analysis. 
The mass distribution of these candidates  
is  shown in  Fig.~\ref{fig_bs_finfit}(a). 
  Two sources of background are identified:  the  combinatorial background and a peaking background component from 
\mbox{$\Lb \rightarrow \phi K^{-} p$}  decays due to  misidentification of  a proton as a kaon.
A maximum-likelihood fit is performed to the mass distribution of selected $\Bs\to\phi\phi$ candidates.  The signal shape is described by the sum of a double-sided Crystal Ball function~\cite{Skwarnicki:1986xj} and a Student's t-function~\cite{Statbook} 
with all parameters fixed to the values estimated in the simulation except the  peak position and resolution. The \mbox{$\Lb \rightarrow \phi K^{-} p$}  background is described by a double-sided Crystal Ball function  with the shape parameters fixed to the values determined using \textcolor{black}{the fast simulation package {\sc RapidSim}~\cite{COWAN2017239}.} 
The combinatorial background is represented by an exponential function.
The yields of the three components, the position and resolution of the signal component and the slope of the background exponential function are allowed to vary in the fit.
The  \mbox{$\Bs\to \phi \phi$} signal yield 
is measured to be $15840\pm140$.  Based on the result of the fit to the mass distribution, a  signal weight 
is assigned to each candidate using the \sPlot method~\cite{Pivk:2004ty}.  These signal weights are subsequently used in a  maximum-likelihood fit~\cite{Xie:2009rka} to the decay-time and angular distributions in order to statistically subtract the background contribution.

 The  decay of a $\Bs$ meson to the \mbox{$K^{+}K^{-}K^{+}K^{-}$} final state can proceed via the $\phi \phi$, $ \phi {f_0}$ and ${f_0}{f_0}$ intermediate states. Due to the small phase space of the decay $f_0 \to \Kp\Km$ and the narrow $\Kp\Km$ mass window used to select the $\phi$ candidates,  the latter two contributions are highly suppressed and found to be  negligible from an angular fit that accounts for these contributions. 
Thus in the subsequent analysis,  only the $\Bs\to\phi \phi$ decay is considered. 
The differential decay rate  is written as  the  sum of six terms, corresponding to contributions from the three polarization states and their interferences,
\begin{equation}
   \label{eqn_timang_rate}
   \frac{{\rm d}^4\Gamma(t,\vec{\Omega})}{ {\rm d}t{\rm d}\vec{\Omega}}
   \propto\sum_{k=1}^{6}h_{k}(t)f_{k}(\vec{\Omega})\;,
\end{equation}
where $t$ is the decay time of the \Bs  meson, and 
$\vec{\Omega}=(\theta_1, \theta_2, \chi)$ denotes the  helicity angles of the two $\Kp$ mesons in the corresponding $\phi$ rest frame ($\theta_1, \theta_2$) and the angle between the  two $\phi\to \Kp\Km$ decay planes ($\chi$).    
The angular functions $f_{k}(\vec{\Omega})$ are defined in Ref.~\cite{LHCb-PAPER-2019-019}. The time-dependent functions
$h_{k}(t)$ are given by\textcolor{black}{ ~\cite{duPree:2010mba}}
\begin{equation}
 \label{eqn_timang_term}
   h_{k}(t)=N_{k} e^{-\Gamma_s{t}} \cbracket{a_{k}\cDGs+
   b_{k}\sDGs 
   +Q  c_{k}\cDms+Q d_{k}\sDms}\;  \nonumber
\end{equation}    
Here $Q$ equals $+1$ ($-1$) for an initial $\Bs$ ($\Bsb$) state, 
$\Delta m_{s}$ is the mass difference between the heavy  and light  \Bs mass 
eigenstates,
$\Delta \Gamma_{s}$ is the decay width difference
between the light and heavy  mass eigenstates, and $\Gamma_{s}$ is the average decay width.
Ignoring  \CP violation in the \Bs mixing, in line with 
experimental measurements~\cite{HFLAV18},
the coefficients $N_k$, $a_k$, $b_k$, $c_k$ and $d_k$ 
are defined~\cite{LHCb-PAPER-2019-019} in terms of  the magnitudes  $\abs{A_i}$,  phases $\delta_i$, \CP-violating phases $\phi_{s,i}$ and direct \CP-violation parameters $\abs{\lambda_i}$ for the three polarization states of the \Bs  decay at $t=0$, with  $i=0,\;\parallel,\;\perp $.
The three amplitudes satisfy 
$\abs{A_0}^2+\abs{A_{\parallel}}^2+\abs{A_{\perp}}^2=1$.
The parameters 
$\phi_{s,i}$ and $\abs{\lambda_i}$
are defined by the equation
\begin{equation}
    \frac{q}{p} \frac{\Bar{A_i}}{A_i} = \eta_i \abs{\lambda_i} e^{-i\phi_{s,i}}\;,
\end{equation}
 where $\eta_i$ is the \CP eigenvalue of the  polarization state $i$,  $q$ and $p$ are complex numbers relating the \Bs mass eigenstates to the flavor eigenstates.  
 A  subset of parameters,
 chosen here as ($\phi_{s,i}$, $\abs{\lambda_i}$, $\abs{A_0}^2$, $\abs{A_{\perp}}^2$, $\delta_{\perp}-\delta_0$, $\delta_{\parallel}-\delta_0$), can be determined   by performing a maximum-likelihood fit to  the distributions of $t$, $\vec{\Omega}$ and $Q$. 
  In the SM-like case or new physics scenarios where  \CP violation is   polarization independent, the set of \CP-violation observables can be reduced to $\phi_{s,i}=\phisss$ and $\abs{\lambda_{i}}=\abs{\lambda}$.    
  In this analysis, the above formalism is used to obtain both polarization-independent and polarization-dependent results, taking into account the  experimental effects discussed below.

The detector acceptance and   selection requirements lead to a nonuniform  efficiency as a function of the angular variables, referred to  below as the angular acceptance. This effect is accounted for through the use of  normalization factors calculated with simulated signal events subject to the  same  
selection criteria as the data. 
 Weights are assigned to the simulated events to improve the agreement with the data, in the shape of the kaon \pt distribution. These weights are determined with an iterative algorithm~\cite{LHCb-PAPER-2013-002, LHCb-PAPER-2019-019}.

The reconstruction, trigger and selection requirements result in  a decay-time dependent efficiency.  
A cubic spline function~\cite{Karbach-2014qba}, with 7 knots at [0.3, 0.5, 1.0, 1.5, 2.0, 3.0, 8.0]~\ps and 9 coefficients, 
is employed  to describe the decay-time dependent efficiency  function, referred to below as the decay-time acceptance.  
One coefficient  is fixed to unity for normalization, and all the other coefficients  are determined in the fit to the data, 
%taking advantage of the known $\Gamma_s$ and $\Delta\Gamma_s$ values.  
\textcolor{black}{
where the parameters $\Gamma_{s}$ and $\Delta\Gamma_{s}$ are constrained to  the recent    measurements by the LHCb collaboration in \mbox{$\Bs\to J/\psi\phi$} decays~\cite{LHCb-PAPER-2019-013}.}
%\mbox{$\Gamma_{s} =0.657 \pm 0.002~\ps^{-1}$} and \mbox{$\Delta\Gamma_{s}= 0.078 \pm 0.006~\ps^{-1}$} with a correlation coefficient of $-0.35$.
Compared with the previous analysis in Ref.~\cite{LHCb-PAPER-2019-019}, which used $\Bs\to D_s^-\pi^+$ and $B^0\to \jpsi K^{*0}$ decays as control channels to determine the decay-time acceptance, this method with free acceptance parameters simplifies the analysis without loss of precision for the  physics parameters.

The dilution effect of the decay-time resolution on the \Bs oscillation  is %effectively
modelled by a  Gaussian
with a 
per-candidate width $\sigma_{t}$, which is related to the per-candidate decay-time uncertainty, $\delta_{t}$, through a linear  calibration function
    $\sigma_t = q_0 + q_1 \times \delta_t$.
The parameters $q_0$ and $q_1$ are obtained  using fictitious  candidates formed of four prompt tracks from $pp$ interactions, which have a decay time centered around 0. These  prompt candidates are weighted to match the  momentum and \pt distributions of the signal candidates and split into ten  $\delta_t$ intervals. 
For each interval, the effective time resolution $\sigma_{t,i}$  is estimated by fitting the sum of  three Gaussian functions with a common mean to 
the decay-time distribution of the prompt candidates and converting the resultant  triple Gaussian into a single Gaussian that has the same damping effect on the observed \Bs oscillation amplitude~\cite{LHCb-PAPER-2019-013}. A linear fit to the data points ($\delta_{t,i}$, $\sigma_{t,i}$) for all intervals provides the estimates of $q_0$ and $q_1$.
The reliability of this calibration method is verified  using simulated samples of signal decays
and prompt candidates. The effective decay-time  resolution  is  found to be between 42 and 44 fs , depending on the data-taking year.

The initial flavor of a \Bs meson is inferred using both  
opposite-side (OS)~\cite{LHCb-PAPER-2011-027} and same-side (SS)~\cite{LHCb-PAPER-2015-056} tagging algorithms. Each of the two methods  yields a tagging  decision $Q$ with a mistag probability $\kappa$ for each \Bs candidate, where $Q =+1,\; -1$ or $0$, if the candidate is classified as \Bs, \Bsb or untagged, respectively. The mistag probability is calibrated using a linear function 
 $\omega = p_0 + p_1 \times \kappa $ for the OS tagging and a quadratic function
$\omega = p_0 + p_1 \times \kappa + p_2\times \kappa^2$ for the SS tagging, where $\omega$ is the corrected mistag probability. The  calibration of the OS mistag probability uses   \mbox{$\B^{+} \to \jpsi \Kp$} decays, for which the  value of $\omega$ in an interval of $\kappa$ can be obtained  from the numbers of correct and wrong decisions. 
The calibration of the SS mistag probability uses flavor-specific $\Bs\to D_s^- \pi^+$ decays, for which the value of $\omega$ in an interval of $\kappa$ is estimated by fitting the decay-time distribution. Detailed descriptions of the calibration procedures can be found in Refs.~\cite{LHCb-PAPER-2013-002,LHCb-PAPER-2019-019}.
The effective tagging efficiency is estimated to be $(5.7\pm 0.5 )$\%, $(6.1\pm 0.7)$\% and $(6.3\pm 0.5)$\% for the  2015-2016, 2017 and 2018 data samples, respectively.
Mistag asymmetries between \Bs and \Bsb decays are evaluated using $B^{\pm}\to J/\psi K^{\pm}$ decays for the OS tagging and prompt $D_s^{\pm}$ decays as a proxy for \Bs and \Bsb decays for the SS tagging, and accounted for in the subsequent signal fit.

 A  weighted 
 maximum-likelihood fit~\cite{Xie:2009rka} is simultaneously performed to the three subsamples of data recorded in 
 2015-2016, 2017 and 2018. The probability density function for each period is based on Eq.~\ref{eqn_timang_rate} and takes into account the effects of angular acceptance, decay-time acceptance, decay--time resolution and mistag probability.  
The parameter $\Delta m_{s}$ is constrained to the measurement by the LHCb collaboration, \mbox{ $\Delta m_{s}=17.766 \pm 0.006~\ps^{-1}$ ~\cite{LHCb-PAPER-2021-005}. } 
%The parameters $\Gamma_{s}$ and $\Delta\Gamma_{s}$ are constrained to  the recent    measurements by the LHCb collaboration in \mbox{$\Bs\to J/\psi\phi$} decays~\cite{LHCb-PAPER-2019-013}: \mbox{$\Gamma_{s} =0.657 \pm 0.002~\ps^{-1}$} and \mbox{$\Delta\Gamma_{s}= 0.078 \pm 0.006~\ps^{-1}$} with a correlation coefficient of $-0.35$. 
  
 The background-subtracted data distributions of the decay time and angular variables with  projections of the polarization-independent fit are shown in Fig.~\ref{fig_bs_finfit}(b-d).   
 The fit results  are summarized in Table~\ref{tab:FinalValue}, which include both statistical and systematic uncertainties. 
 The correlation matrix is given in  the Supplemental Material~\cite{RefSup}, as well as the observed time-dependent asymmetry between flavor-tagged  \Bs and \Bsb decays.

A summary of the systematic uncertainties for the polarization-independent fit  is reported  in Table~\ref{tab:syst_sum}. The total systematic uncertainty on \phisss, which is  the sum in quadrature of the different contributions, is 0.009 rad, significantly 
smaller than the statistical uncertainty of 0.075 rad. This is also the case for the other physics parameters.

Inaccuracies in the determination of the calibration parameters of the decay-time resolution ($q_0$, $q_1$)  and the flavor tagging ($p_0$, $p_1$, $p_2$) 
lead to  systematic effects on the \CP-violation observables. %For both cases, 
These parameters are fixed in the baseline fit, and their combined statistical and systematic  uncertainties are propagated  to the fit parameters  as systematic uncertainties. The compatibility  of the calibration parameter values in control and signal channels are checked using  simulated samples, and the observed differences are treated as sources of systematic uncertainties. These uncertainties are taken to be fully correlated between different data-taking periods. A  bias 
of roughly $-5$ $\fs$  on the reconstructed decay time is observed in the calibration of the decay-time resolution. \textcolor{black}{Since this bias is much smaller than the effective resolution, neglecting it in the analysis of the signal decays results in negligible  effects on the physics parameters}.

The systematic uncertainties related to the angular
acceptance comprise the  uncertainties due to the limited size of the simulation sample, evaluated using a bootstrapping~\cite{Efron1986BootstrapMF} procedure, and uncertainties associated with convergence of the iterative correction procedure,  estimated by comparing fit results with different numbers of iterations.

The systematic uncertainties related to the decay-time acceptance are found by varying the knot positions of the spline function to achieve similar fit quality of the decay-time distribution to that of the baseline fit. The maximum changes of the parameter estimates are assigned as systematic uncertainties. 

The effects of mismodelling the signal and background mass distributions are studied by using alternative shape functions that can achieve similar fit quality to that of the baseline fit. The maximum changes of the parameter estimates are assigned as systematic uncertainties. 
The \sPlot method used to subtract background in the time-dependent angular fit requires that the discriminating variable, $m(\Kp\Km\Kp\Km)$, is uncorrelated with the decay angles, $\vec{\Omega}$, 
and decay time, $t$.
To evaluate the effects of possible correlations among these variables,   
the data sample is split into three subsamples according to the value of
$\cos^2 \theta_1 + \cos^2 \theta_2$, $\chi$ or  $t$, respectively. A fit to the $m(\Kp\Km\Kp\Km)$ distribution of each subsample is performed, and signal weights are calculated accordingly. The three subsamples are then recombined and a new time-dependent angular fit is performed. The maximum changes of the parameter estimates are  assigned as systematic uncertainties.

\begin{table}[tb]
\caption{
\label{tab:FinalValue} 
    Measured observables in the polarization-independent fit.  
    The first uncertainties are statistical and the second  systematic.
}
\begin{center}
\begin{tabular}{lccc}
   \hline 
Parameter&  Result \\
\hline
	   $\phisss$ [\rad]              &  $-0.042\pm 0.075 \pm0.009 $\\
	   $|\lambda|$  &                  
          $ \;\;\;1.004\pm0.030\pm0.009$\\
	   
	   $\abs{\Azero}^2$   &            
         $\;\;\;0.384 \pm0.007\pm0.003$\\
         $\abs{\Aperp}^2$ &              
          $\;\;\;0.310\pm0.006 \pm0.003$\\
          $\delpara-\delzero$ \;[\rad]  &  
         $ \;\;\;2.463\pm0.029 \pm0.009$\\
	   $\delperp-\delzero$ \;[\rad]&    
          $ \;\;\;2.769\pm0.105\pm0.011$\\  
	 
   \hline
\end{tabular}
\end{center}
\end{table}

\begin{table}[tb]
\begin{center}
        \caption{Systematic uncertainties  for physics parameters
        in the polarization-independent fit, the values are given in units of $10^{-3}$ ($10^{-3}$ rad for angles). 
}
        \label{tab:syst_sum}
        \begin{tabular}{lccccccc}
      
          \hline 
Source&$\phisss\;  $&$|\lambda|$&$|A_0|^2$  &$|A_\perp|^2$ & $\delpara-\delzero\;$ & $\delperp-\delzero\;$  \\
         \hline
         Time resolution&4.9&2.6&0.8&0.8&0.1&3.4\\
         Flavor tagging &4.8&4.7 &0.9&1.3&1.2&9.7&\\
         Angular acceptance&3.9&4.9&1.4&1.7&4.7&1.2\\
         Time acceptance &2.3&1.7&0.1&0.1&5.6&0.7\\
         Mass fit \& factorization &2.2&4.4 & 1.9&2.3&2.3&2.5\\
          MC truth match&1.1&0.2 &0.1&0.1&0.2&0.3\\
         Fit bias&0.8&0.7&0.9&0.3&3.6&0.7\\
         Candidate multiplicity&0.3&0.2 &0.1&0.8&0.2&0.1\\
        
\hline
         Total &8.8&8.6 &2.7&3.3&8.5&10.7\\
          \hline 
        \end{tabular}
\end{center}
\end{table}

\begin{figure}[b]
\begin{center}
\includegraphics[width=0.60\linewidth]{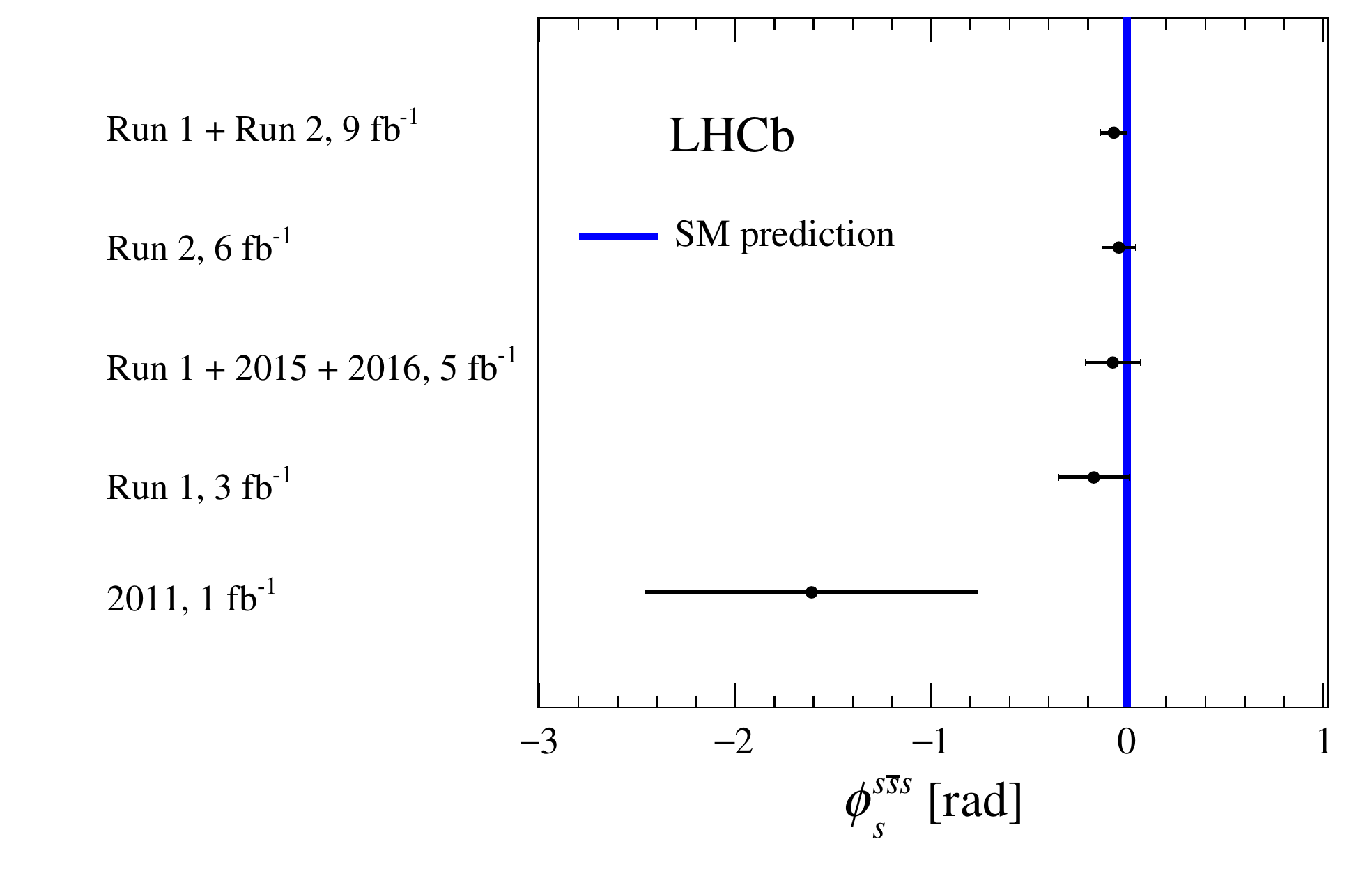}\\
 \end{center}
 \vspace*{-1cm}
 \caption{ 
    Comparison of  \phisss measurements from this and previous analyses~\cite{LHCb-PAPER-2013-007, LHCb-PAPER-2014-026, LHCb-PAPER-2019-019}  
     by the  \lhcb  collaboration.  
     The vertical band indicates the SM prediction~\cite{Raidal:2002ph, Beneke:2006hg, Cheng:2009mu}.
   }
\label{fig_LHCb_measurement}

\end{figure}

The intrinsic bias from the maximum-likelihood fit with a limited sample size is evaluated by performing pseudoexperiments, which also demonstrate that the parameter uncertainties estimated in the fit are reliable after correcting for the background dilution effect.

Various checks of the fit procedure are performed by splitting the data sample according to  magnet polarity, trigger selection, tagging category, data-taking period, and multiple decay-time  and \Bs-meson \pt  intervals. The effect of tightening  the kaon-identification and MLP-output requirements is also  studied. The fit results are compatible between  different subsamples in all checks.

The polarization-independent  measurements of the \CP-violation parameters  
$\phisss$ and $\abs{\lambda}$ in \mbox{$\Bs\to\phi\phi$} decays presented here are combined with the LHCb Run 1 
measurements, $\phisss = -0.17 \pm 0.15 \stat \pm 0.03 \syst$\rad and 
 \mbox{$\abs{\lambda}=1.04\pm 0.07 \stat \pm 0.03 \syst$~\cite{LHCb-PAPER-2014-026}}
 using the procedure described in Ref.~\cite{LHCb-PAPER-2019-013}. 
 In the combination,  
those systematic uncertainties that arise from the same origin  
are taken to be completely correlated between the Run~1 and Run~2 results.    
The combined values of the \CP-violation parameters are 
$\phisss = -0.074 \pm 0.069$\rad 
and  $\abs{\lambda} = 1.009 \pm 0.030 $, with a correlation coefficient of $-0.02$. This is the most precise measurement of \CP violation in $\Bs \to \phi \phi$ decays to date, as is illustrated in Fig.~\ref{fig_LHCb_measurement}.

A polarization-dependent fit is  performed using the same data set, where the parameters $\phi_{s,i}$ and 
 $\lambda_{i}$  can take different values for the three polarization states. To reduce parameter correlations in the fit, the phase differences, $\phipa -\phi_{s,0}$ and $\phipe -\phi_{s,0}$, and ratios, $|\lamperp/\lamzero|$ and  $|\lampara/\lamzero|$, are used as fit parameters. 
 The measured values are

\begin{align}
\phi_{s,0}  &= - 0.18\pm 0.09 \; \rad \;,&       |\lambda_{0}| =   1.02 \pm 0.17 \;,  \nonumber \\
\phipa -\phi_{s,0} &= \;\;\; 0.12 \pm 0.09 \; \rad \;, &    |\lamperp/\lamzero|  = 0.97 \pm  0.22 \;,  \nonumber \\
\phipe -\phi_{s,0} &= \;\;\; 0.17 \pm 0.09 \; \rad \;, &     |\lampara/\lamzero| = 0.78 \pm 0.21 \;,\nonumber  
\end{align}
where the uncertainties are statistical only.
No significant difference   between 
different polarization states is observed.

In conclusion,  a measurement of the polarization-independent \CP-violation observables  in  \mbox{$\Bs\to\phi\phi$} decays    is performed using  data  collected with the LHCb detector in 2015--2018,  corresponding to a total integrated luminosity of  6\invfb. The results are 
\begin{eqnarray}
\begin{split}
\phisss & = -0.042 \pm 0.075 \pm 0.009 \, \rad \, , 
 \nonumber \\
\abs{\lambda} & = \;\;\;1.004\pm 0.030 \pm 0.009 \,,  \nonumber
\end{split}
\end{eqnarray}
where the first uncertainties are statistical and the second systematic. 
These results are combined with the LHCb measurements based on data taken in 2011 and 2012  to obtain 
$\phisss = -0.074 \pm 0.069$\rad and  
$\abs{\lambda} = 1.009 \pm 0.030$.
This is the most precise measurement of time-dependent \CP asymmetry in the decay $\Bs\to\phi\phi$  and in any penguin-dominated $B$ meson decay. The measurement is consistent with and supersedes the  measurement in Ref.~\cite{LHCb-PAPER-2019-019}, and agrees with the SM expectation of tiny \CP violation. 
For the first time, the polarization-dependent \CP-violation parameters are measured, which  show no significant difference between the three polarization states of \mbox{$\Bs\to\phi\phi$}  decays.
These results can be used to constrain new physics contributions in $b\to s$ transitions~\cite{Zhang:2000ic, Datta:2012ky, Bhattacharya:2013sga}.

\section*{Acknowledgements}
%
% These Acknowledgements valid from 3-May-2019
%
\noindent We express our gratitude to our colleagues in the CERN
accelerator departments for the excellent performance of the LHC. We
thank the technical and administrative staff at the LHCb
institutes.
We acknowledge support from CERN and from the national agencies:
CAPES, CNPq, FAPERJ and FINEP (Brazil); 
MOST and NSFC (China); 
CNRS/IN2P3 (France); 
BMBF, DFG and MPG (Germany); 
INFN (Italy); 
NWO (Netherlands); 
MNiSW and NCN (Poland); 
MCID/IFA (Romania); 
%MSHE (Russia); 
MICINN (Spain); 
SNSF and SER (Switzerland); 
NASU (Ukraine); 
STFC (United Kingdom); 
DOE NP and NSF (USA).
We acknowledge the computing resources that are provided by CERN, IN2P3
(France), KIT and DESY (Germany), INFN (Italy), SURF (Netherlands),
PIC (Spain), GridPP (United Kingdom), 
%RRCKI and Yandex LLC (Russia), 
CSCS (Switzerland), IFIN-HH (Romania), CBPF (Brazil),
Polish WLCG  (Poland) and NERSC (USA).
We are indebted to the communities behind the multiple open-source
software packages on which we depend.
Individual groups or members have received support from
ARC and ARDC (Australia);
Minciencias (Colombia);
AvH Foundation (Germany);
EPLANET, Marie Sk\l{}odowska-Curie Actions, ERC and NextGenerationEU (European Union);
A*MIDEX, ANR, IPhU and Labex P2IO, and R\'{e}gion Auvergne-Rh\^{o}ne-Alpes (France);
Key Research Program of Frontier Sciences of CAS, CAS PIFI, CAS CCEPP, 
Fundamental Research Funds for the Central Universities, 
and Sci. \& Tech. Program of Guangzhou (China);
%Key Research Program of Frontier Sciences of CAS, CAS PIFI,
%Thousand Talents Program, and Sci. \& Tech. Program of Guangzhou (China);
%RFBR, RSF and Yandex LLC (Russia);
GVA, XuntaGal, GENCAT, Inditex, InTalent and Prog.~Atracci\'on Talento, CM (Spain);
SRC (Sweden);
the Leverhulme Trust, the Royal Society
 and UKRI (United Kingdom).

\clearpage
\ifx\mcitethebibliography\mciteundefinedmacro
\PackageError{LHCb.bst}{mciteplus.sty has not been loaded}
{This bibstyle requires the use of the mciteplus package.}\fi
\providecommand{\href}[2]{#2}

% LHCb collaboration author list
% Data extracted on April 19th, 2023 at 11:13am for paper reference LHCb-PAPER-2023-001
\centerline
{\large\bf LHCb collaboration}
\begin
{flushleft}
\small
R.~Aaij$^{32}$\lhcborcid{0000-0003-0533-1952},
A.S.W.~Abdelmotteleb$^{51}$\lhcborcid{0000-0001-7905-0542},
C.~Abellan~Beteta$^{45}$,
F.~Abudin{\'e}n$^{51}$\lhcborcid{0000-0002-6737-3528},
T.~Ackernley$^{55}$\lhcborcid{0000-0002-5951-3498},
B.~Adeva$^{41}$\lhcborcid{0000-0001-9756-3712},
M.~Adinolfi$^{49}$\lhcborcid{0000-0002-1326-1264},
P.~Adlarson$^{77}$\lhcborcid{0000-0001-6280-3851},
H.~Afsharnia$^{9}$,
C.~Agapopoulou$^{43}$\lhcborcid{0000-0002-2368-0147},
C.A.~Aidala$^{78}$\lhcborcid{0000-0001-9540-4988}
% ...

{\footnotesize \it

% ...
$^{9}$Universit{\'e} Clermont Auvergne, CNRS/IN2P3, LPC, Clermont-Ferrand, France\\
% ...
$^{32}$Nikhef National Institute for Subatomic Physics, Amsterdam, Netherlands\\
% ...
$^{41}$Instituto Galego de F{\'\i}sica de Altas Enerx{\'\i}as (IGFAE), Universidade de Santiago de Compostela, Santiago de Compostela, Spain\\
% ...
$^{43}$European Organization for Nuclear Research (CERN), Geneva, Switzerland\\
% ...
$^{45}$Physik-Institut, Universit{\"a}t Z{\"u}rich, Z{\"u}rich, Switzerland\\
% ...
$^{49}$H.H. Wills Physics Laboratory, University of Bristol, Bristol, United Kingdom\\
% ...
$^{51}$Department of Physics, University of Warwick, Coventry, United Kingdom\\
% ...
$^{54}$School of Physics and Astronomy, University of Glasgow, Glasgow, United Kingdom\\
$^{55}$Oliver Lodge Laboratory, University of Liverpool, Liverpool, United Kingdom\\
% ...
$^{63}$Syracuse University, Syracuse, NY, United States\\
% ...
$^{77}$Department of Physics and Astronomy, Uppsala University, Uppsala, Sweden, associated to $^{54}$\\
$^{78}$University of Michigan, Ann Arbor, MI, United States, associated to $^{63}$\\
\bigskip
% ...
}
\end{flushleft}

\end{document}